\documentclass{ws-procs975x65}
\usepackage[english]{babel}

\begin{document}

\title{Gauge fields and particle-like formations associated with shear-free null congruences
}
\author{Vladimir V.~Kassandrov and Vladimir N.~Trishin
}
\address{Department of General Physics, Peoples' Friendship University of Russia, \\
Ordjonikidze 3, 117419, Moscow, Russia\\
E-mail: vkassan@sci.pfu.edu.ru}

\maketitle

{\it Twistors} encode the space-time background geometry and, on the
other hand, relate to the structure of null geodesic congruences (NGC).
Particularly, the {\it Kerr theorem}~\cite{penrose} in twistor terms describes
the complete set of {\it shear-free} NGC of the flat space: it asserts
that any 2-spinor $\xi_{A'}$ being a solution of the equation
\begin{equation}\label{inv}
\Pi(\xi_{A'}, iX^{AA'}\xi_{A'}) = 0, ~~~~(A,A',...=0,1)
\end{equation}
(and, consequently, a function of the space-time coordinates $X^{AA'}$)
gives rise to a shear-free NGC with tangent vector $k_\mu =\xi_A\xi_{A'}$.
And that, moreover, any (analytical) shear-free NGC may be obtained via this
construction. Here $\Pi$ is an arbitrary {\it homogeneous} (analytical) functions
of three complex variables -- components of the {\it projective} null twistor
$W=\{\xi, iX\xi\}$ of the Minkowski space.

Shear-free NGC constitute one of the two classes of solutions to complex
eikonal equation~\cite{eik} with respect to the (ambi)twistor structure of the
latter. Besides, any (analytical) congruence defines the
strength of a null Maxwell field~\cite{robin} and relates to the solutions of
free massless equations~\cite{penrose}.  Defining equations of shear-free NGC
follow from (\ref{inv}) and have the form
\begin{equation}\label{SFC}
\xi^{A'}\xi^{B'}\partial_{AA'}\xi_{B'}=0.
\end{equation}
These equations appear also in the framework of
{\it noncommutative analysis} as the {\it conditions of differentiability} of
functions taking values in the algebra of complex quaternions (biquaternions)
$\mathbb{B}$. Remarkably, only on the base of eqs.(\ref{SFC}) (or of their
general solution represented by the Kerr constraint (\ref{inv})) a
self-consistent nonlinear field theory has been developed~\cite{AD,GR,acta}.
Correspondent approach has been therein called {\it algebrodynamics}.

Specifically, in the framework of the analysis over $\mathbb{B}$-algebra one comes to the
following invariant matrix differentiability conditions ({\it generalized Cauchy-Riemann
equations})~\cite{AD,GR,acta,prot,jos}:
\begin{equation}\label{GSE}
d\xi = \Phi dX \xi, ~~~\Leftrightarrow ~~~\partial_{AA'} \xi_{B'}=\Phi_{B'A}\xi_{A'}
\end{equation}
After elimination of the auxiliary complex field $\Phi_{A'A}$ eqs.(\ref{GSE})
reduce to the {\it over-determined} system of four equations
\begin{equation}\label{GCRE}
\xi^{A'}\partial_{AA'}\xi_{B'}=0 ,
\end{equation}
from which, evidently, the shear-free condition (\ref{SFC}) does follow.
Moreover, for the scaling-invariant {\it ratio} of spinor components
eqs. (\ref{SFC}) and (\ref{GCRE}) turn to be equivalent, and thus
{\it for any shear-free congruence some field $\Phi(X)$ can be defined}
(up to a gauge transformation, see below). Thus, the structures
represented by eqs. (\ref{inv}),(\ref{SFC}),(\ref{GSE}), ({\ref{GCRE}) are
in fact equivalent.

Geometrically, eqs.(\ref{GSE}) can be thought of as conditions for the spinor field
$\xi$ to be {\it parallel} with respect to effective affine connection $\Omega=\Phi dX$.
In the 4-vector representation it gives rise to exceptional geometry with Weyl
nonmetricity and skew symmetric torsion~\cite{AD,GR,jos}. By this, both the Weyl 4-vector
and the torsion pseudotrace are proportional to each other and to complex field
$\Phi_{A'A}$.

Following Weyl's idea, the quantities $\Phi_{A'A}$ can be identified as the
components of 4-potential of a complex gauge field. Indeed, {\it integrability
conditions} of eqs. (\ref{GSE}) are precisely the conditions of {\it self-duality}
of matrix connection $\Omega$. From this and the Bianchi identities it follows
that {\it complex Maxwell and $SL(2,\mathbb{C})$ Yang-Mills equations are satisfied
on the solutions of (\ref{GSE})}, for the trace and the trace-free parts of
connection $\Omega$ respectively~\cite{GR,jos}.

The gauge group for these fields is, however, {\it restricted} (as well as
the residual rescaling group of the shear-free equations in the form (\ref{GCRE})).
Specifically, only the following ``weak'' gauge transformations preserve the
form of systems (\ref{GSE}) and (\ref{GCRE})~\cite{GR,jos}:
\begin{equation}\label{gauge}
\xi_{A'}\mapsto\hat\xi_{A'}=\alpha\xi_{A'} \\
~~~~~\phi_{A'A}\mapsto\hat\phi_{A'A}=\phi_{A'A}+\partial_{A'A} \ln \alpha,
\end{equation}
where the gauge parameter $\alpha$ can't depend on the space-time coordinates explicitly
but {\it only through the components of the transforming twistor $W$} (i.e. of the spinor
$\xi$ and its counterpart $\tau=iX\xi$).

Correspondent field strengths allow for explicit representation through the (second
order) derivatives of the spinor field or, still more, for a nice representation in
twistor variables~\cite{jos,eik}. But the most interesting feature of this field is,
perhaps, the  property of {\it quantization of electric charge} of any bounded field
singularity, in particular of Coulomb-like type. Theorem of quantization which
establishes also the  existence of {\it minimal ``elementary'' charge} has been proved
in~\cite{vest,sing} and is based on the over-determinancy of shear-free equations and on the
self-duality of complex field strength (for more details about the concept of singular
sources of Maxwell field and the charge quantization see~\cite{trish,sing}).

On these grounds, one can identify (bounded) singularities of effective
Maxwell field with {\it particle-like} formations. With respect to the initial
shear-free NGC they are nothing but the {\it caustics} of the congruence and,
in view of the generating Kerr constraint (\ref{inv}), defined by the condition
\begin{equation}\label{sing}
\frac{d\Pi}{dG}=0
\end{equation}
($G$ being the {\it ratio} of spinor components on which the homogeneous
function $\Pi$ depends essentially). For a given $\Pi$,
one can eliminate the projective spinor $G$ from the joint system of eqs.
(\ref{inv}) and (\ref{sing}) and come to a constraint on coordinates $f(X)=0$.
This defines (at a fixed moment of time) {\it the shape of singular locus} of
the congruence and the associated fields and governs also their time evolution.
The function $f(X)$ obeys the complex eikonal equation~\cite{time}. A number of interesting
examples of quite nontrivial distributions and dynamics of singularities
(which can be point-, string- or even 2D-membrane-like and undergo various
bifurcations simulating transmutations and even annihilations of particles)
has been studied in our works~\cite{trish,jos,prot,sing,time}.
By all, the above presented
algebraic method to obtain enormously complicated singular solutions
of Maxwell and Yang-Mills equations (or, at least, the structure of their
singular loci) seems very powerful and has, perhaps, no analogues
in literature.

There exists a simple generalization of the scheme to the curved space.
In fact, a Riemannian metric of the {\it Kerr-Schild type}
\begin{equation}\label{schild}
g_{\mu\nu} = \eta_{\mu\nu} + H k_\mu k_\nu
\end{equation}
may be defined for any starting shear-free NGC $k_\mu=\xi_A\xi_{A'}$ in Minkowski space
$\eta_{\mu\nu}$ ($H=H(X)$ being a scalar function of coordinates). Remarkably,
the properties of the congruence to be null, geodesic and shear-free remain invariant
under such deformations of space-time geometry, and curvature singularities of metrics
(\ref{schild}) coincide with those of the electromagnetic field strengths $F_{AB}$ being
represented by the same condition (\ref{sing}). If the ray of the congruence is a principal
direction of $F_{AB}$ which satisfies thus the constraint $F_{AB}\xi^A\xi^B = 0$,
then the electromagnetic field remains self-dual in the Kerr-Schild space
(\ref{schild}) \cite{TC98}. Note that precisely such Maxwell fields (generally 
independent from those above considered) have been explicitly constructed for 
shear-free congruences in \cite{LN,NL}. Together with correspondent metric 
(and under the proper choice of the factor $H(X)$) they can satisfy, moreover, 
the Einstein-Maxwell electrovacuum system. 

In particular, this is the case of 
the Kerr-Newman stationary solution~\cite{schild}
generated by the {\it Kerr congruence} with the twofold structure and the
singular ring - caustic. Remarkably, this ring
(as the source of electromagnetic and gravitational fields) possesses
the value of gyromagnetic ratio inherent for Dirac's fermion, and 
B.~Carter~\cite{carter}, A.~Ya.~Burinskii~\cite{burin} and others 
even considered it as a possible model of electron. In the framework 
of our algebrodynamical scheme this is still more natural since the electric charge 
is here necessarily {\it unit}, ``elementary'' in modulus. 
E.~T.~Newman\cite{NP69} proposed to interpret "matter" as elementary singularities and 
systematically studied them in the framework of complex space-time 
representation. 

More successively, one has to study the shear-free NGC directly on the Riemannian
background~\cite{GRG}. For a wide class of {\it special} shear-free congruences (SSFC)
the tangent vector $k_\mu$ is a {\it repeated} principal direction of the Weyl curvature
tensor. SSFC preserve many remarkable properties of shear-free NGC of the flat space. In
particular, for any SSFC its equations, under the proper gauge, reduce to the form
(\ref{GCRE}) (with evident account of the Levi-Civita connection,
$\partial\leftrightarrow\nabla$). The gauge field $\Phi$ can also be defined in a similar
manner, and the SSFC spinor $\xi$ is again parallel with respect to  effective
Weyl-Cartan connection of the above presented type.  However, the self-duality conditions
are generally broken in the presence of curvature term, and therefore, an effective
geometrical source appear in the Maxwell equations for the associated field $\Phi$. In
asymptotically flat spaces the presence of this source doesn't damage the
Coulombian asymptotics and thus the quantization rule for admissible values of electric
charge is inherited from the flat space.

To conclude, the algebrodynamical approach based on shear-free NGC equations or on
related twistor and biquaternionic structures leads to a peculiar self-consistent
dynamics of associated fields and to complicated particle-like distributions of the
congruence fields which manifest nontrivial interactions and quantum-like properties. The
approach leads also to new conjectures about the {\it nature of physical time}, the {\it
light-born matter} and to naturally {\it multivalued} structure of primordial field
entities (see~\cite{time} and references therein).

The authors thank E. T. Newman for sending his new paper \cite{NL} on similar 
issues.

\end{document}